\documentclass[twocolumn, tighten]{aastex631}

\usepackage{amsmath}

\newcommand{\Oxford}{\affiliation{Department of Physics, University of Oxford, 
Parks Road, Oxford OX1 3PU, UK}}

\usepackage{float}
\usepackage{bm}

\newcommand{\KIAA}{\affiliation{Kavli Institute for Astronomy and
Astrophysics, Peking University, Beijing 100871, China}}

\newcommand{\DOA}{\affiliation{Department of Astronomy, School of Physics,
Peking University, Beijing 100871, China}}

\newcommand{\NAOC}{\affiliation{National Astronomical Observatories,
Chinese Academy of Sciences, Beijing 100012, China}}

\begin{document}

\title{Impact of overlapping signals on parameterized post-Newtonian coefficients in tests of gravity }

\correspondingauthor{Ziming Wang}
\email{zwang@pku.edu.cn}

\correspondingauthor{Lijing Shao}
\email{lshao@pku.edu.cn}

\author[0000-0002-6689-8680]{Yixuan Dang}\Oxford  
\author[0000-0002-8742-8397]{Ziming Wang}\DOA\KIAA
\author[0000-0001-5021-235X]{Dicong Liang}\KIAA
\author[0000-0002-1334-8853]{Lijing Shao}\KIAA\NAOC

\begin{abstract}

Gravitational waves have been instrumental in providing deep insights into the nature of gravity. Next-generation detectors, such as the Einstein Telescope, are predicted to have a higher detection rate given the increased sensitivity and lower cut-off frequency. However, this increased sensitivity raises challenges concerning parameter estimation due to the foreseeable overlap of signals from multiple sources. 
Overlapping signals (OSs), if not properly identified, may introduce biases in estimating post-Newtonian (PN) coefficients in parameterized tests of general relativity (GR). We investigate how OSs affect $-1$PN to 2PN terms in parameterized GR tests, examining their potential to falsely suggest GR deviations. 
We estimate the prevalence of such misleading signals in next-generation detectors, and their collective influence on GR tests. 
We compare the effects of OSs on coefficients at different PN orders, concluding that overall the 1PN coefficient suffers the most. 
Our findings also reveal that while a non-negligible portion of OSs exhibit biases in PN coefficients that might individually prefer to conclude deviations from GR, collectively, the direction to deviate is random and a statistical combination will still be in favor of GR.
\end{abstract}


\section{Introduction} \label{sec:intro}
Gravitational waves (GWs), predicted by general relativity (GR), have provided essential tests of gravity since their first direct detection by the LIGO/Virgo Collaboration in 2015~\citep{LIGOScientific:2016aoc}. Current detected GWs are all generated from compact binary coalescences, such as the mergers of binary black holes (BBHs) or neutron stars, and carry information about the sources and the nature of gravity itself. 
To date, observatories including LIGO, Virgo, and KAGRA have identified approximately 90 GW events, offering us some of the most profound insights into GR and significantly enhancing our grasp of gravitational dynamics in the strong-field regime~\citep{LIGOScientific:2017ycc,LIGOScientific:2017bnn,LIGOScientific:2018mvr,LIGOScientific:2021djp}.
Various tests of GR were also performed with current detections with no conclusive sign of GR being falsified~\citep{LIGOScientific:2018dkp,LIGOScientific:2019fpa,LIGOScientific:2020tif,LIGOScientific:2021sio, ghosh2022summary}.

Next generation (XG) detectors, for example, the Einstein Telescope~\citep[ET;][]{Punturo:2010zz,Hild:2010id,Sathyaprakash:2012jk} and the Cosmic Explorer~\citep[CE;][]{Reitze:2019iox,Reitze:2019dyk}, are expected to receive far more signals than current detectors do due to the largely increased sensitivity and a lower cut-off frequency~\citep{Sathyaprakash:2019yqt, Kalogera:2021bya}. In particular, ET, whose sensitivity is expected to be better by a factor of 10 than current-generation ground based detectors, has been shown to detect overlapping signals (OSs) in the mock data challenges~\citep{Regimbau:2012ir,Meacher:2015rex}. 

While the enhanced sensitivity of detectors is a significant achievement, it introduces a new challenge in the parameter estimation (PE) process that several studies have explored, notably in dealing with the potential overlap of signals from multiple GW sources~\citep{Samajdar:2021egv,Pizzati:2021apa,Relton:2021cax,Regimbau:2009rk}.
Such OSs can complicate the PE process, introducing biases across parameters~\citep{Antonelli:2021vwg, Janquart:2022nyz, Wang:2023ldq}. In this work we investigate if misidentifying OSs can distort PE on parameters in the post-Newtonian (PN) waveform~\citep{Yunes:2009ke}, suggesting deviations from GR. In the case where a weaker secondary signal overlaps with a primary signal, erroneously classifying this as a single, pure signal can lead to inaccuracies. It leads us to three crucial questions:
\begin{itemize}
    \item Is it possible for OSs, if inaccurately perceived as from a singular source, to misguide us into concluding a deviation from GR?
    \item Are such deceptive OSs frequent enough to pose a genuine concern for XG detectors? 
    \item If a substantial number of such signals are identified, do they, as a group, indicate a deviation from GR? 
\end{itemize}
There exist abundant studies on tests of parameterized PN coefficients for GR in general~\citep{Arun:2006hn, Mishra:2010tp, Cornish:2011ys, LIGOScientific:2021sio,Wang:2022yxb}. For OSs, \cite{Hu:2022bji} have shown that they may lead to incorrect measurement of deviation from the 0PN term in the GR waveform.

However, a comprehensive assessment of the effects of OSs on PN coefficients is lacking. 
Our work aims to explore the aforementioned questions related to OSs regarding false-alarmed deviations in various PN coefficients, including $-1$PN, 0PN, 0.5PN, 1PN, 1.5PN, 2PN coefficients, and offers a comparative analysis of how each PN coefficient responds to misidentified OSs. 
We anticipate not only a potential for misleading deviations across all PN terms but also that some PN terms might be more adversely affected  than others when leaving out the secondary signals in the OSs. 

Our results show that there exist OSs resulting in large bias in multiple PN coefficients, and notably, these misleading signals represent a non-negligible portion of all overlapping instances to be registered by XG detectors. 
We also find that the relative size of the fractions of misleading OSs at different PN orders depends on how we define OSs. If only two GWs with very small merger-time difference can be identified as OSs, the fraction of misleading OSs at higher PN orders is larger. 
But generally, the impact of overlapping on the $1$PN term is most significant among $-1$PN to 2PN coefficients, that is, GR-violation effects are most likely to be observed at 1PN when misidentifying an OS as one single GW signal.
Fortunately, these signals only prefer models with deviation from GR individually, but not collectively, manifested in the rapid decrease of Bayes factor with increased occurrences of misleading signals. This aligns with the result in \cite{Hu:2022bji}, which considered the deviation from the 0PN term when OSs exist.

This paper is organized as follows. We introduce PE methods used in this work in Sec.~\ref{sec:methods}, including how we calculate the Fisher matrix, biases, and Bayes factor for OSs.
In Sec.~\ref{sec:sim setup} we list our simulation setup for both the waveforms and the population model for BBHs. Our results are given in Sec.~\ref{sec:result}, and Sec.~\ref{sec:con and dis} concludes the work.


\section{Parameter estimation methods} \label{sec:methods}


To determine the parameters $\bm{\theta}$ of the GW signal from data $g(t)$ received by GW detectors, one can calculate the posterior distribution of the parameters using Bayes' theorem,
\begin{equation}
    P(\boldsymbol{\theta}|g(t))\propto P(\boldsymbol{\theta})P(g|\boldsymbol{\theta}),
    \label{bayes theorem}
\end{equation}
where $P(\boldsymbol{\theta})$ is the prior of the parameters $\boldsymbol{\theta}$ given assumed background knowledge, and $P(g|\boldsymbol{\theta})$ is the conditional probability of receiving data $g(t)$ given parameters $\boldsymbol{\theta}$. 
The data $g(t)$ consist of the GW signal $h(t)$ and noise $n(t)$.
Assuming the noise is stationary and Gaussian, the likelihood reads~\citep{Finn:1992wt}
\begin{equation}
    P(g|\boldsymbol{\theta}) \propto e^{-\frac{1}{2}(g-h,g-h)},
    \label{likelihood}
\end{equation}
where the inner product $(u,v)$ is defined as
\begin{equation}
    (u,v): = 2\Re\int_{-\infty}^{\infty}\frac{u^{*}(f)v(f)}{S_n(|f|)} { d} f,
\end{equation}
where $S_n$ is the power spectral density of the noise, $u(f)$
and $v(f)$ are the Fourier transforms of $u(t)$ and $v(t)$, respectively.

In principle, given the GW data, one can use some
sampling techniques, such as the Markov-Chain Monte Carlo (MCMC) method~\citep{Christensen:1998gf, Christensen:2004jm, Sharma:2017wfu} and Nested
Sampling~\citep{2004AIPC..735..395S, Skilling:2006gxv}, to obtain the posterior
distributions of $\bm \theta$. 
However, the full Bayesian inference takes considerable computational time and resources, especially when the dimension of parameter space  is high. 
In this work, we adopt the Fisher matrix (FM) as a fast approximation of the full Bayesian inference, which has been successfully applied to conducting PE for numerous OSs in previous works~\citep{Antonelli:2021vwg,Hu:2022bji,Wang:2023ldq}. 
Mathematically, the FM is defined as~\citep{Cutler:1994ys,Vallisneri:2007ev}
\begin{equation}
    F_{\alpha \beta}=(h_{,\alpha},h_{,\beta})\,,
\end{equation}
where $h_{,\alpha}\equiv \partial h/\partial \theta^\alpha$. Under the high signal-to-noise ratio (SNR) limit, or equivalently the linear signal approximation (LSA), the posterior can be well approximated by a Gaussian distribution~\citep{Finn:1992wt,Vallisneri:2007ev}. Denoting the inverse of the FM as $\Sigma \equiv F^{-1}$ and assuming flat priors, the standard deviations and correlation coefficients of parameters are given by (no summation over repeated indices)
\begin{equation}\label{FM approximation}
  \sigma_{\alpha}=\sqrt{\Sigma^{\alpha\alpha}}\,, \quad c_{\alpha\beta}=\frac{\Sigma^{\alpha\beta}}{\sqrt{\Sigma^{\alpha\alpha}\Sigma^{\beta\beta}}}\,.
\end{equation}


Considering a superposition of two independent GW signals in the detector, the data can be written as 
\begin{equation}
	g(t)= h^{(1)}(t;\tilde{\bm \theta}^{(1)}) +h^{(2)}(t;\tilde{\bm \theta}^{(2)})+n(t) \,,
\end{equation}
where $\tilde{\bm \theta}^{(i)}$ represents the true parameters of the $i$-th signal. 
We consider the situation where an OS is mistakenly categorized as a  signal containing a single event, and therefore only one-signal model $h(t;{\bm \theta})$ is used in the PE. 
Usually, the louder signal can be identified, while the weaker one induces biases in the PE of the louder signal ~\citep{Pizzati:2021apa}.
Without loss of generality, we assume that signal 1 is louder than signal 2. 

When conducting PE, it is more convenient to consider the ratio of the bias relative to the statistical uncertainty of the parameters, that is, the reduced bias~\citep{Wang:2023ldq} 
\begin{equation}
    B_\alpha:=\frac{{\Delta \theta}^\alpha_{\rm bias}}{{\sigma_\alpha}}\, ,
    \label{eq:bias}
\end{equation}
where ${\Delta \theta}^\alpha_{\rm{bias}}={\Sigma^{\alpha \beta}\big(h_{, \beta}^{(1)}, h^{(2)}\big)}$ refers to the absolute bias. Note that the index $\alpha$ is not summed. In this formula, we have assumed  LSA, meaning that the SNR of signal 1 is large enough. 

In  PE, when $|B_{\alpha}|$ exceeds 1, the bias is significant.
 When $\theta^\alpha$ is chosen as a PN-deviation parameter in the parameterized PN waveform test, we call the OSs with $|B_{\alpha}|>1$ as ``misleading OSs'', which means that this kind of OSs may lead to a false conclusion of a deviation from GR. 


Bayes factor is another useful indicator for assessing whether the data show a stronger preference for GR or for a deviation from GR. The Bayes factor $K_{\mathrm{M_1},\mathrm{M_0}}$ is defined as the ratio of evidences,
\begin{equation}
    K_{\mathrm{M_1},\mathrm{M_0}}=\frac{P(D|\mathrm{M_1})}{P(D|\mathrm{M_0})},
    \label{bayes factor}
\end{equation}
where $D$ stands for data, and $\mathrm{M_1}$, $\mathrm{M_0}$ are two competing models. 
Bayes factor compares how well each model predicts the data.
Roughly speaking, when $K_{\mathrm{M_1},\mathrm{M_0}}$ tends to an extreme large/small value, the data favor/disfavor $\mathrm{M_1}$ over $\mathrm{M_0}$. Otherwise, the preference for $\mathrm{M_1}$ and $\mathrm{M_0}$ is comparable.


In our calculation, $D$ is the GW signal $g(t)$, $\mathrm{M_1}$ represents a model allowing the waveform deviating from GR at $-1$, 0, 1 and 2 PN orders, respectively denoted as $\rm{M}_{-1\mathrm{PN}}$, $\rm{M}_{0\mathrm{PN}}$, $\rm{M}_{1\mathrm{PN}}$ and $\rm{M}_{2\mathrm{PN}}$.
The model $\mathrm{M_0}$ represents GR. 
As an example, assuming flat priors and LSA, the Bayes factor of $\rm{M}_{-1\mathrm{PN}}$ against ${\rm {M}_0}$ is given by~\citep[also see Eq. (13) in ][]{Hu:2022bji}
\begin{equation}
    \begin{aligned}
    K_{\rm{M}_{-1\mathrm{PN}},\rm{M}_0} &= \frac{P \big(g(t)|\rm{M}_{-1\mathrm{PN}} \big)}{P \big(g(t)|\rm{M}_0 \big)} \\
    &= p \big( {\delta\phi_{-1}  } \big) \sqrt{2 \pi} \frac{e^{- \frac{1}{2} \hat{\chi}_{\rm{M}_{-1\mathrm{PN}}}^2}}{e^{- \frac{1}{2} \hat{\chi}_{\rm{M}_0}^2 }} \sqrt{  \frac{|\Sigma_{\rm{M}_{-1\mathrm{PN}}}|}{| \Sigma_{\rm{M}_0}|}},
    \end{aligned}
    \label{eq:bayes factor specific}
\end{equation}
where $p \big( {\delta\phi_{-1}  } \big)$ is the prior probability for the $-1$PN deviation parameter $\delta\phi_{-1}$. We incline to consider $p \big( {\delta\phi_{-1}  } \big)$ as a constant. The residual chi-square, $\hat{\chi}^2_{{\rm M}_i}$, is defined as
\begin{equation}
 \hat{\chi}^2_{{\rm M}_i}=\left(h^{(2)}-h_{, \alpha}^{(1)} \, \Delta{\theta}^\alpha_{{\rm bias,M}_i}, h^{(2)}-h_{, \alpha}^{(1)} \, \Delta{\theta}^\alpha_{{\rm bias,M}_i}\right),
\end{equation}
where $\Delta{\theta}^\alpha_{{\rm bias,M}_i}$ is the absolute bias of parameter $\alpha$ under hypothesis ${\rm M}_i$.

\section{Simulation setup} \label{sec:sim setup}


We generate non-spinning, stellar-mass BBH signals using the {\sc{IMRPhenomD}} waveform model. 
Similarly to \cite{Wang:2023ldq}, we average the waveform over the sky position, inclination and polarization angles. 
This corresponds to a conservative scenario, where two GW signals come from the same direction with the same inclination angle and polarization angle. 
Using the angle-averaged waveform will affect the fraction of misleading OSs. 
However, here we are more interested in comparing the fractions of misleading OSs for different PN deviations, which are hardly affected by the angle-averaging. 

When generating OSs, we use the GR waveform for both signals 1 and 2, and the free parameters for each signal are the chirp mass $\mathcal{M}$ in the detector frame, the symmetry mass ratio $\eta$, the luminosity distance $d_L$, and the time of coalescence $t_c$. When we recover with one single GR waveform, the estimated parameters are simply $\{{\cal M},\eta,d_L,t_c\}$ of the primary signal.  

For the GR-deviation model at the $i$-th PN order, we add a deviation parameter $\delta \phi_{i}$, which leads to a phase correction in the Fourier-domain waveform~\citep{LIGOScientific:2021sio}
\begin{equation}
    \Delta \Phi_i(f)=\frac{3}{128\eta} \mathfrak{u}^{2i-5}\delta \phi_i \,,
\end{equation}
where $\mathfrak{u} \equiv (\pi Mf)^{1/3}$, and $M$ is the redshifted total mass. 
We only consider the modification in the inspiral stage, which cuts off when the frequency is twice the innermost stable circular orbit frequency. Usually, the deviation parameters at different PN orders are studied separately, assuming only one PN deviation parameter is nonzero, which is more effective to pick up possible deviations at different PN orders~\citep{LIGOScientific:2021sio}. Therefore, for the $i$-th PN deviation model, only 5 parameters, $\{{\cal M},\eta,d_L,t_c,\delta \phi_i\}$, are estimated.

As for the detector, we choose the ET as an example, which is expected to detect thousands of OSs per year~\citep{Pizzati:2021apa}. We will use the designed ET-D sensitivity curve~\citep{Punturo:2010zz,Hild:2010id,Maggiore:2019uih}.



Now we introduce the population model. When generating OSs, we inject signal 1 and signal 2 with GR waveform, which requires the binary masses, the luminosity distance and the time of coalescence of the two signals. We also request the SNR of signal 1 to be larger than 30 to ensure the validity of FM approximation. For the secondary signal, which is randomly drawn from population samples, we require its SNR to be smaller than that of the primary. Our treatment here is somehow arbitrary, but reasonable. More sophisticated treatments might be adopted in future studies.

The BBH mass population is generated according to the POWER+PEAK phenomenological model given in \cite{LIGOScientific:2018jsj,KAGRA:2021duu}.
For the distribution of the luminosity distance, we adopt the model given in \cite{Regimbau:2016ike}, which gives the merger rate in the detector frame,
\begin{equation}
R_z(z)=\frac{R_m(z)}{1+z} \frac{d V(z)}{d z} \,,
\end{equation}
where $V$ is the comoving volume. $R_m(z)$ is the merger rate per comoving volume at redshift $z$,
\begin{equation}
    R_{m}(z) = \int_{t_{\min }}^{t_{\max }}  R_{f}\left(\tilde{z}\left[\tilde{t}(z)-t_d\right]\right) P\left(t_{d}\right)  d t_{d} \,.
    \label{eq:merger rate}
\end{equation}
Here, $t_d$ is time delay between the formation of a binary and its merger, $\tilde{t}(z)$ is the cosmic time when the merger happens, $\tilde{z}$ is the redshift as a function of cosmic time,
$P(t_d)$ is the distribution of the time delay, which is inversely proportional to $t_d$ and normalized between $t_{\min}$ and $t_{\max}$. Following previous works, we choose $t_d^{\mathrm{min}} = 50\,\mathrm{Myr}$~\citep{Meacher:2015iua,LIGOScientific:2017zlf}, and $t_d^{\mathrm{max}}$ equal to the Hubble time~\citep{ Ando:2004pc,Belczynski:2006br,Nakar:2007yr,OShaughnessy:2007brt,Dominik:2012kk,Dominik:2013tma}. The star formation rate  $R_{f}\left(z\right)$ is given by
\begin{equation}
    R_{f}(z)=\nu \frac{a e^{b\left(z-z_{p}\right)}}{a-b+b e^{a\left(z-z_{p}\right)}} \,,
\end{equation}
where $\nu=0.146 \,\mathrm{M}_\odot \, \mathrm{yr}^{-1} \, \mathrm{Mpc}^{-3}$, $a=2.80$, $b=2.46$ and $z_p=1.72$~\citep{Vangioni:2014axa,Behroozi:2014tna}.

The population model of $t_c$ needs more discussion. On the one hand, both the biases and the Bayes factor only depend on the difference between $t_c^{(1)}$ and $t_c^{(2)}$, denoted as $\Delta t_c$, so we only need to consider the distribution of $\Delta t_c$. 
On the other hand, in this work we are interested in the fraction of misleading OSs in all OSs, but there is no unified criterion for OSs~\citep{Wang:2023ldq}. 
Usually, this is done by setting a threshold $\Delta t_{c,\rm{th}}$ for $\Delta t_c$ between two signals, which ranges from 0.1\,s to several seconds in the previous works~\citep{Pizzati:2021apa,Relton:2021cax,Himemoto:2021ukb}. 
To avoid the ambiguity of choosing $\Delta t_{c,\rm th}$, we vary $\Delta t_{c,\rm{th}}$ in a wide range, from 0.1\,s to ${\cal{O}}(10)$\,s and generate the corresponding $t_c$ samples for each threshold. 
Consider an observation with a duration of $T$, and two GW events that are equally likely to occur at any time in this observation, i.e., both $t_c^{(1)}$ and $t_c^{(2)}$ obey a uniform distribution in the time interval $(0, T)$. 
We find that the conditional distribution of $\Delta t_c$ given a threshold, $P\Big(t_c^{(2)}-t_c^{(1)}\Big| |t_c^{(2)}-t_c^{(1)}|\leq \Delta t_{c,\rm th} \Big)$, can be well approximated by a uniform distribution. 




\section{Results} \label{sec:result}


\begin{figure*}    
    \centering   \includegraphics[width=0.9\textwidth]{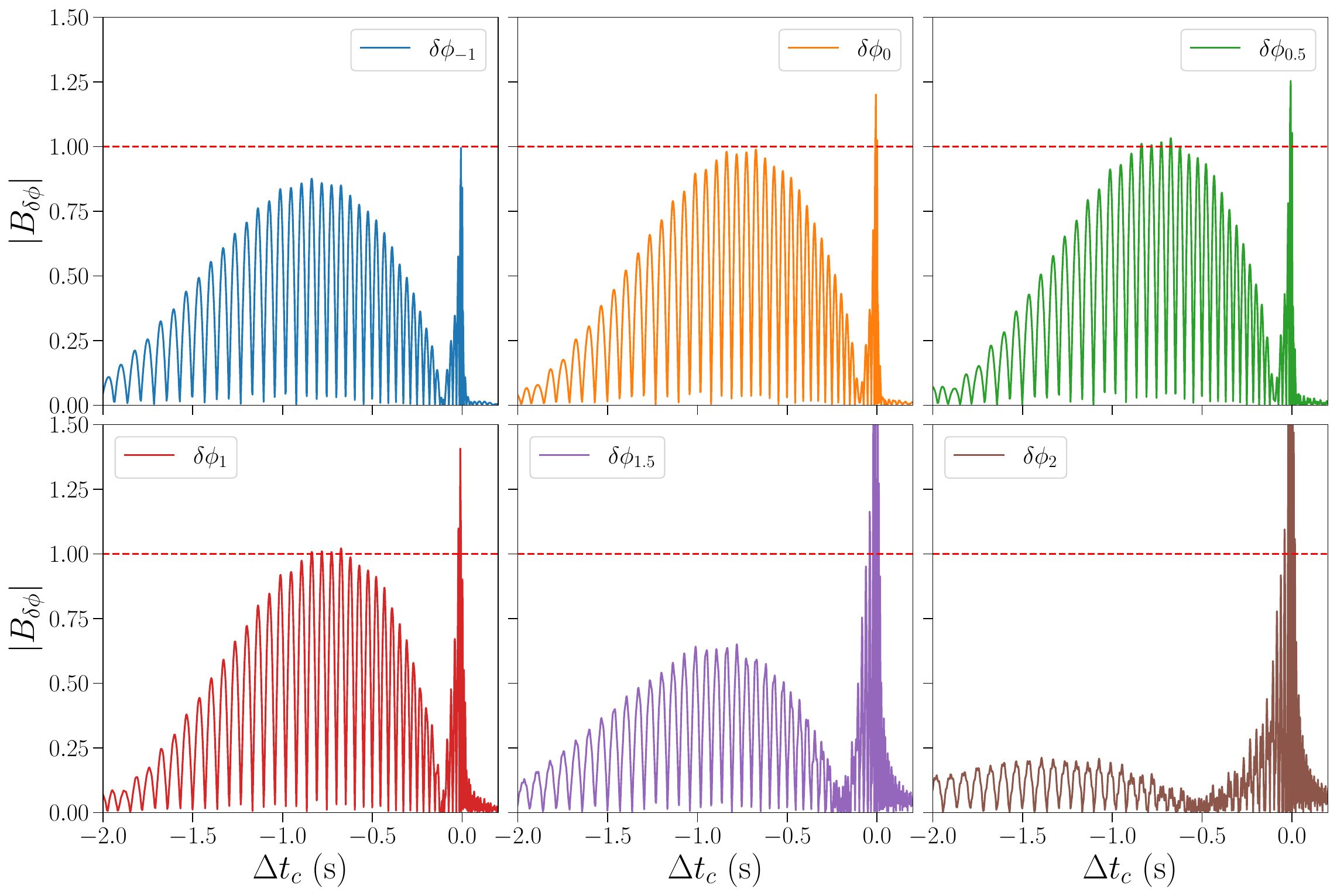}
    \caption{Reduced bias of each PN deviation term as a function of $\Delta t_c$, for an OS characterized by  parameters in Eq.~(\ref{eq:OS:example:pars}).} 
    \label{fig:PNs_Bias_tc}
\end{figure*}

\begin{figure}    
    \centering   
    \includegraphics[width=0.45\textwidth]{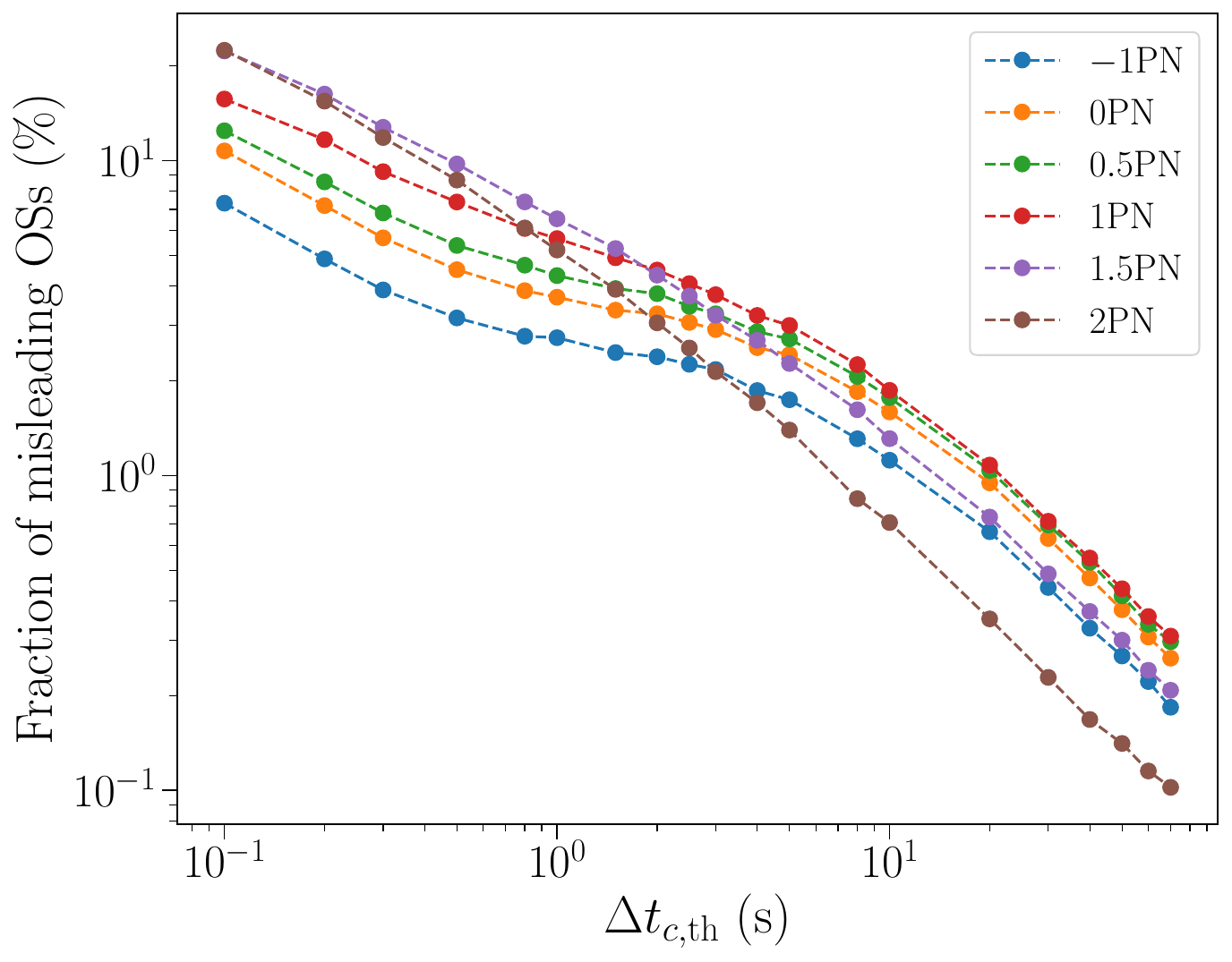}
    \caption{Fraction of misleading OSs for each PN-deviation model against $\Delta t_{c,\rm th}$.}
    \label{fig:PNs_tcs}
\end{figure}

As an illustration, we consider an OS with the following parameters,
\begin{align}
	\mathcal{M}^{(1)} &= 32\,\mathrm{M}_\odot \,,  \quad &
	\mathcal{M}^{(2)} &= 35\,\mathrm{M}_\odot \,,  \nonumber \\
	\eta^{(1)} &=0.246\,,  \quad &
	\eta^{(2)} &=0.248\,,  \nonumber \\ 
	d_{L}^{(1)} &=10000\,\rm{Mpc}\,,  \quad & 
	d_{L}^{(2)} &=45000\,\rm{Mpc}\,,  \nonumber \\ 
	\rm{SNR}^{(1)} &=41.74\,,  \quad &
	\rm{SNR}^{(2)} &=9.96\,. \label{eq:OS:example:pars}
\end{align}
In Figure~\ref{fig:PNs_Bias_tc} we plot the change of  the reduced bias with respect to $\Delta t_c$ for this  OS. It answers our first question in Sec.~\ref{sec:intro}, namely, it is possible that, under the effects of a weaker signal 2, some PN terms of signal 1 possess a reduced bias larger than 1. 
It is also worth attention that large biases of $-1$PN, 0PN, 0.5PN, 1PN (and less significantly 1.5PN) coefficients are obtained around $\Delta t_c=1 \,{\rm s}$, while the bias of $2$PN coefficient has peaked at a much smaller $\Delta t_c$ around zero. 
This can be explained by the fact that the higher order PN terms are mainly prominent in the late inspiral phase of the signal, and we expect the deviation from a higher order PN term to be more significant when $\Delta t_c$ between two signals is small.

Simply highlighting a single instance with parameters in Eq.~(\ref{eq:OS:example:pars}) is not sufficient to underscore the risks posed by all OSs, as such instances may be too infrequent to a warrant general concern. 
To answer the second question in Sec.~\ref{sec:intro}, we generate populations for different $\Delta t_{c,\rm th}$ and calculate the biases for every PN-deviation waveform under consideration. 
Afterwards, we calculated the fraction of misleading OSs in all OSs in each scenario, where the event numbers are large enough to ignore the statistical fluctuations. The findings are presented in Figure~\ref{fig:PNs_tcs}.
The observation is that the fractions of misleading OSs, defined as $|B_{\delta \phi}| > 1$, are around ten percent for populations with small $\Delta t_{c,\rm th}$. Such fractions are not negligible. 
All fractions then decrease as $\Delta t_{c,\rm th}$ increases, because most signals with strong overlapping features happen at small $\Delta t_c$. 
When $\Delta t_{c,\rm th}$ is large, e.g., larger than 10\,s, all fractions are inversely proportional to $\Delta t_{c,\rm th}$. This means that almost all misleading OSs have $\Delta t_c \leq 10\,{\rm s}$.\footnote{The fraction of the misleading OSs is 
$$\frac{N({\rm misleading\,\, OSs})}{N({\rm all\,\, OSs})}=\frac{N({\rm misleading \,\, OSs})}{N({\rm OSs}|\Delta t_c \leq 10{\rm s})}\cdot \frac{N({\rm OSs}|\Delta t_c \leq 10{\rm s})}{N({\rm all\,\,OSs})} \,.$$ If all misleading OSs have $\Delta t_c \leq 10\,{\rm s}$, the first factor on the right-hand side of the above equation does not depend on $\Delta t_{c,\rm th}$, while the second factor is inversely proportional to $\Delta t_{c,\rm th}$. }

The behaviors of these fractions in Figure~\ref{fig:PNs_tcs} at different PN orders are also noteworthy. We note that the fraction of misleading OSs in the 2PN-deviation and 1.5PN-deviation cases is larger than that in the 1PN-deviation case when $\Delta t_{c,\rm th}$ is within $0.5$\,s, while for a larger $\Delta t_{c,\rm th}$ it is the opposite. 
Since the 2PN and 1.5PN terms are more prominent in the late inspiral phase, it is more likely to generate misleading OSs when $\Delta t_c$ is small (as shown in Figure~\ref{fig:PNs_Bias_tc}), corresponding to a relatively large fraction. 
As a trade-off, almost all misleading OSs in the 2PN- and 1.5PN-deviation cases have $\Delta t_c \lesssim 0.5\,{\rm s}$, corresponding to the quick decrease after $\Delta t_{c, {\rm th}} \simeq 0.5\,$s. 
In comparison, in the 0PN, 0.5PN and 1PN deviation cases, similar behavior of these fractions is observed at $\Delta t_{c,\rm th}$ between 1\,s and 10\,s. The fraction of misleading OSs in the 1PN-deviation case decreases faster than that at the 0PN and 0.5PN orders, meaning that the number of misleading OSs in the 0PN- and 0.5PN-deviation cases with $1\,{\rm s}<\Delta t_c<10\,{\rm s}$ is larger than that in the 1PN-deviation case. 
However, since $\Delta t_c$ is relatively large when the 0PN and 0.5PN term dominates the phase evolution, it is difficult to generate large biases, and the total number of misleading OSs in the 1PN-deviation case is still larger than that in the 0PN-deviation case for a large $\Delta t_{c,\rm th}$. In general, for most $\Delta t_{c,\rm th}$ (from 2\,s to 70\,s), the fraction of large bias in 1PN coefficient stays the highest among the PN-deviation cases we investigate in this work. 


\begin{figure*}    
    \centering   \includegraphics[width=0.95\textwidth]{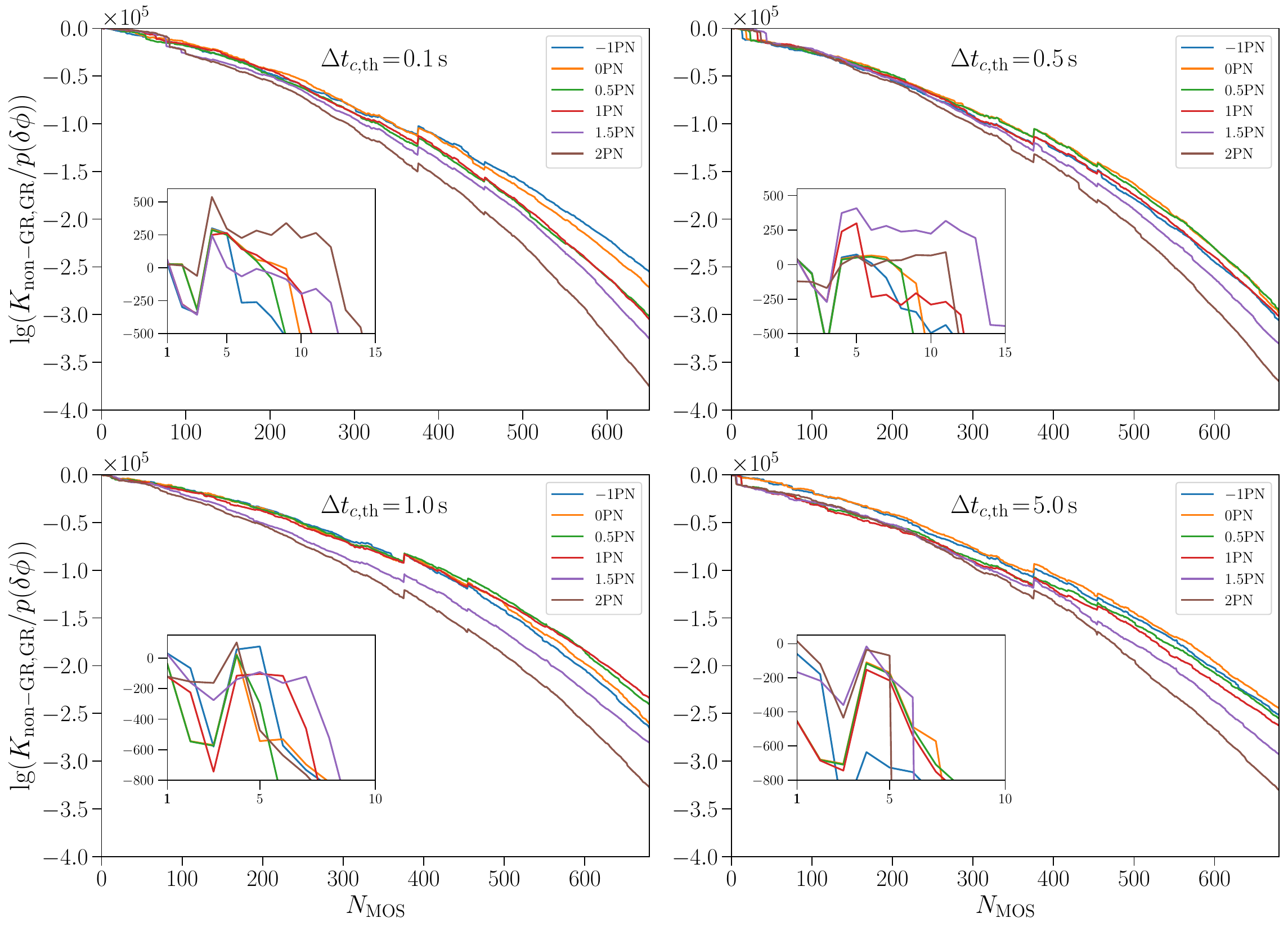}
    \caption{Bayes factor divided by $p(\delta \phi)$ over accumulation of misleading OSs, denoted as $N_{\mathrm{MOS}}$, with different $\Delta t_{c,\rm th}$ for each panel. The inset within each plot is a zoom-in plot of the initial behavior of the Bayes factor.}
    \label{fig:Kf}
\end{figure*}

From the above findings, we observe a non-negligible proportion of misleading OSs to parameterized PN waveform tests. These OSs, when analyzed individually in tests of GR, tend to favor their respective PN-deviation model more strongly relative to the GR waveform. However, this is not the case when all misleading OSs are considered collectively because of the disagreement on the values of deviation parameters arising from each misleading OS. Figure~\ref{fig:Kf} shows the evolution of Bayes factors of each PN-deviation model against GR with the increase of number of misleading OSs, denoted as $N_\mathrm{MOS}$. 
For all PN-deviation models, their corresponding Bayes factors exceed 1 only for the first few misleading OSs, and then start drastically decreasing as more misleading OSs enter the test of GR.
It is evident that biases in PN deviation terms caused by OSs are not constant as they do not represent a ``real'' deviation. This is because these biases arise solely from the misclassification of an OS as a single-event signal. 

\begin{figure*}    
    \centering   \includegraphics[width=0.95\textwidth]{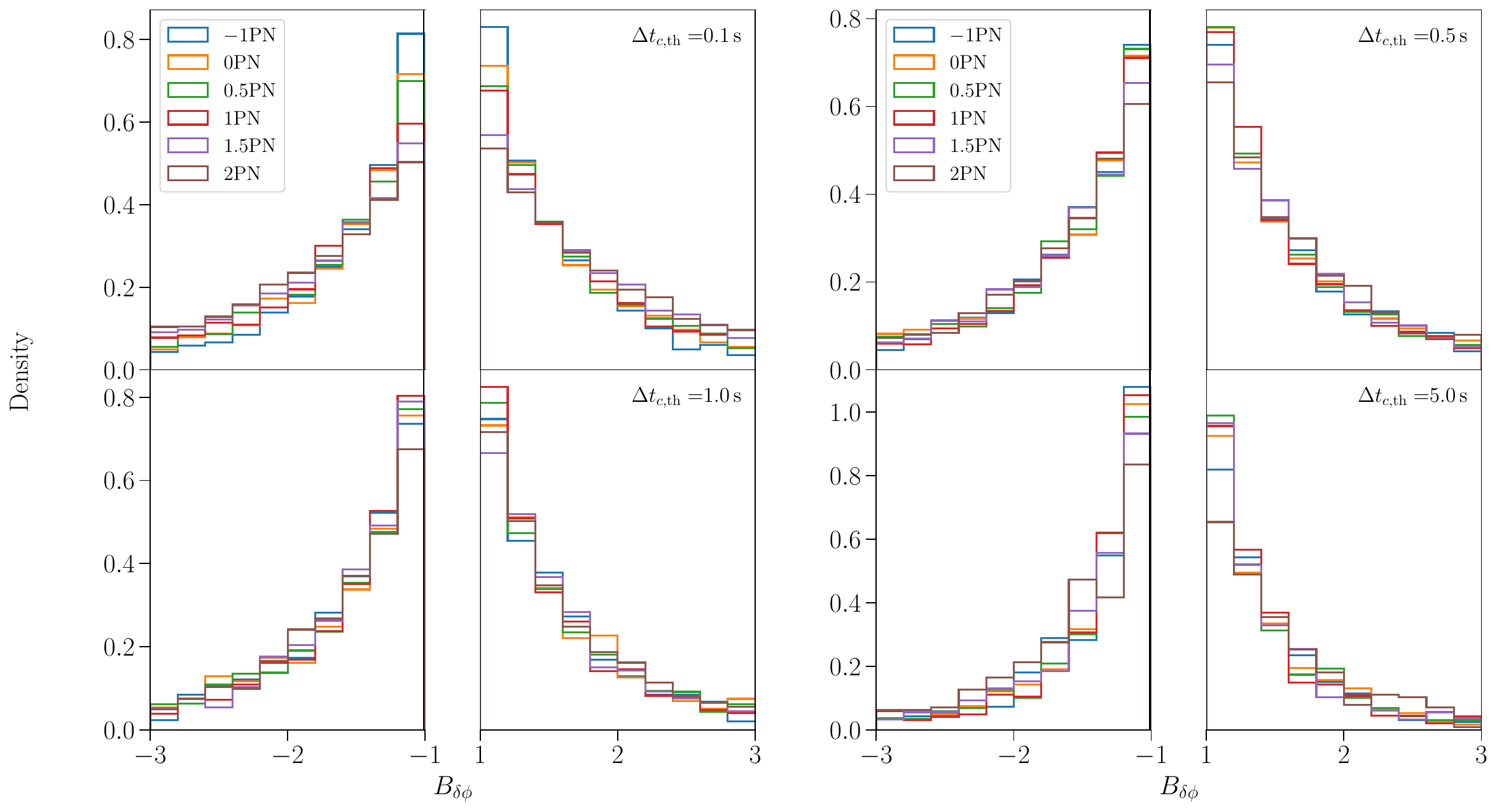}
    \caption{Distribution of large biases of each PN deviation parameter for different $\Delta t_{c,\rm th}$.}
    \label{fig:PNs_bias_distri}
\end{figure*}

However, in our PN-deviation models, the deviation parameters are assumed to be constant. When combining two misleading OSs, where, say, one has the reduced bias $B_\alpha>1$ and the other has $B_\alpha<-1$, the Bayes factor will rapidly decrease due to the inconsistent biases. To support this, we plot the distribution of reduced biases of the misleading OSs for each PN deviation model in Figure~\ref{fig:PNs_bias_distri}, from which we can observe symmetric distributions of reduced biases. 

Another interesting aspect noted in Figure~\ref{fig:Kf} is that the Bayes factor for the 2PN and 1.5PN deviation models decline more rapidly than those for the $-1$PN, 0PN, 0.5PN, and 1PN deviation models, especially at small $\Delta t_{c,\rm th}$.
In Figure~\ref{fig:PNs_bias_distri}, one can find that the distributions of biases of the $-1$PN, 0PN, 0.5PN and 1PN deviation models cluster more densely near 1, while the distributions for the 2PN and 1.5PN deviation models are broader, leading to a greater variance in individual biased values. 
This is consistent with the argument above that it is more likely to have larger biases in the deviation parameter at the 2PN and 1.5PN orders. 
Therefore, the Bayes factor for the 2PN and 1.5PN deviation models decrease more rapidly with accumulation of misleading OSs.

\section{Conclusions and Discussions} \label{sec:con and dis}

The XG ground-based detectors are anticipated to exhibit a significantly enhanced detection rate of BBHs relative to their forerunners, indicating the unavoidable appearance of OSs in future detection~\citep{Regimbau:2009rk, Meacher:2015rex,Kalogera:2021bya}.
A primary concern regarding such OSs is how they can potentially introduce biases if an OS is mistakenly treated as a single-event signal, thereby misleading parameterized tests of GR. 
\cite{Hu:2022bji} studied the biases in PE when testing GR at the 0PN order with these OSs, and left the comprehensive investigation at different PN orders for future work.


In this work, we extend \citet{Hu:2022bji} and investigate how OSs affect $-1$PN to 2PN terms in parameterized tests of GR. We adopt the FM approximation to calculate the biases introduced by OSs. FM is a fast and effective method to study the biases of OSs~\citep{Finn:1992wt,Antonelli:2021vwg,Hu:2022bji,Wang:2023ldq}.
For each PN deviation parameter, we calculate the proportion of signals exhibiting significant bias, and further explore the Bayes factor over accumulation of such misleading OSs. We find that a non-negligible fraction of OSs could produce a significant bias in each PN deviation parameter, especially when the difference in coalescence times, $\Delta t_c$, is small. It is worth noting that there are significant differences between the fractions of misleading OSs at different PN orders, while the fraction in the 1PN-deviation case usually stays the largest.
This underscores the importance of consideration for GR deviations at different PN orders when discussing the impact of OSs on testing gravity.

However, different misleading OSs lead to inconsistent values for deviation parameters, and if analyzed collectively, they do not show a preference for non-GR deviations over GR. 
This is supported by the decreasing trend of the Bayes factor with an increasing number of misleading OSs, shown in Figure~\ref{fig:Kf}. 

In real detection, there exist other sources of PE biases when analyzing OSs, such as the waveform modeling~\citep{Hu:2022bji}, instrumental calibration~\citep{Sun:2020wke,Hall:2017off} and detector glitches~\citep{Pankow:2018qpo}. We leave a detailed assessment and comparison of all these effects on testing gravity for future work. However, as the misleading OSs discussed, we emphasize the importance of considering the effects at different PN orders to obtain a global understanding.




\section*{Acknowledgements}

We thank the anonymous referee for useful comments.
This work was supported by the Beijing Municipal Natural Science Foundation (1242018), the National Natural Science Foundation of China
(11975027, 11991053), the China Postdoctoral Science Foundation (2021TQ0018), the National SKA Program of China
(2020SKA0120300), the Roy Skinner fund from Balliol College at University of Oxford,
the Max Planck Partner Group Program funded by the Max Planck Society, 
and the High-Performance Computing Platform of Peking University. 
\facilities{Einstein Telescope}

\bibliographystyle{aasjournal}  
\bibliography{main}  
\end{document}